\documentclass[12pt,preprint]{aastex}

\shorttitle{Variability in GLIMPSE I \& MSX}
\shortauthors{Robitaille et al.}

\newcommand{\microns}{$\mu$m~}
\newcommand{\micronsns}{$\mu$m}

\newcommand{\irac}{IRAC 8.0\micronsns~}
\newcommand{\msx}{MSX 8.28\micronsns~}
\newcommand{\mips}{MIPS 24.0\micronsns~}
\newcommand{\msxe}{MSX 21.3\micronsns~}
\newcommand{\iracns}{IRAC 8.0\micronsns}
\newcommand{\msxns}{MSX 8.28\micronsns}
\newcommand{\mipsns}{MIPS 24.0\micronsns}
\newcommand{\msxens}{MSX 21.3\micronsns}
\newcommand{\ks}{K$_{\rm s}$~}
\newcommand{\ksns}{K$_{\rm s}$}
\newcommand{\av}{A$_{\rm V}$}

\begin{document}

\title{Infrared point source variability between the Spitzer and MSX surveys of the Galactic mid-plane}

\author{Thomas P. Robitaille\altaffilmark{1}, Martin Cohen\altaffilmark{2}, Barbara A. Whitney\altaffilmark{3}, Marilyn Meade\altaffilmark{4}, Brian Babler\altaffilmark{4}, Remy Indebetouw\altaffilmark{5}, Ed Churchwell\altaffilmark{4}}

\altaffiltext{1}{SUPA, School of Physics and Astronomy, University of St Andrews, North Haugh, KY16 9SS, St Andrews, United Kingdom; tr9@st-andrews.ac.uk}
\altaffiltext{2}{Radio Astronomy Laboratory, 601 Campbell Hall, University of California at Berkeley, Berkeley, CA 94720; mcohen@astro.berkeley.edu}

\altaffiltext{3}{Space Science Institute, 4750 Walnut St. Suite 205, Boulder, CO 80301, USA; bwhitney@spacescience.org}
\altaffiltext{4}{Department of Astronomy, 475 North Charter St., University of Wisconsin, Madison, WI 53706; meade@astro.wisc.edu, brian@sal.wisc.edu, ebc@astro.wisc.edu}
\altaffiltext{5}{Spitzer Fellow, University of Virginia, Astronomy Dept., P.O. Box 3818, Charlottesville, VA, 22903-0818; remy@virginia.edu}

\begin{abstract}
We present a list of 552 sources with suspected variability, based on a comparison of mid-infrared photometry from the GLIMPSE~I and MSX surveys, which were carried out nearly a decade apart.
We were careful to address issues such as the difference in resolution and sensitivity between the two surveys, as well as the differences in the spectral responses of the instruments.
We selected only sources where the \irac and \msx fluxes differ by more than a factor of two, in order to minimize contamination from sources where the difference in fluxes at 8\microns is due to a strong 10\microns silicate feature.
We present a subset of 40 sources for which additional evidence suggests variability, using 2MASS and MIPSGAL data.
Based on a comparison with the variability flags in the IRAS and MSX Point-Source Catalogs we estimate that at least a quarter of the 552 sources, and at least half of the 40 sources are truly variable.
In addition, we tentatively confirm the variability of one source using multi-epoch IRAS LRS spectra.
We suggest that most of the sources in our list are likely to be Asymptotic Giant Branch stars.
\end{abstract}


%\maketitle

\section{Introduction}

The Combined General Catalog of Variable Stars (GCVS4.2 ; \citealt{samus04}) lists 38,624 confirmed variable sources, most of which are located in the Milky-Way Galaxy. The types of variable sources are numerous, ranging from eruptive (e.g. FU Orionis objects) or pulsating (e.g. Cepheid, Mira, or RR Lyrae) variables to eclipsing binaries or cataclysmic variables. Most of these have been found through optical surveys, which are limited in depth by dust extinction in our own Galaxy. Thus, looking in the direction of the Galactic plane, these sources are likely to be within a kpc from the sun.

Searching for variability at mid-IR wavelengths has two advantages. First, the much reduced extinction at mid-IR wavelengths means that bright stars can be seen to much larger distances. Second, dust emission is much brighter at mid-IR wavelengths than at shorter wavelengths, and therefore we can search for variability in dusty objects such as Asymptotic Giant Branch (AGB) stars or Young Stellar Objects (YSOs).

Data from the IRAS satellite were used to assign variability likelihoods to sources in the IRAS Point Source Catalog (the \texttt{Var} index). This was made possible by the fact that most areas of the sky were observed at least twice, with observations typically separated by weeks to months. Thousands of new variable stars were discovered, most of which lie in the Galactic plane or in the Galactic bulge \citep{spear93}. However, the low sensitivity and resolution of the IRAS observations along with source confusion meant that only the brightest sources were found in the Galactic mid-plane (most sources in the IRAS Point-Source Catalog with $|b|<1^\circ$ and a \texttt{Var} index larger than 98\% - indicating very likely variability - have F$_{12\mu{\rm m}}>1$Jy).

The \emph{Midcourse Space Experiment} (MSX) satellite, launched in 1996, completed a survey of the Galactic plane \citep{price01} from 8 to 21\micronsns, with improved sensitivity and spatial resolution over IRAS. Most areas of the Galactic plane were observed multiple times (typically four times) within a few months. Each source in the MSX Point Source Catalog was then assigned a variability flag in each band (i.e. \texttt{vara}, \texttt{varc}, \texttt{vard}, and \texttt{vare}).

More recently, the Galactic plane was surveyed in the mid-IR by the {\it Spitzer Space Telescope} \citep{werner04}, launched in 2003. {\it Spitzer} carried out the Galactic Legacy Infrared Mid-Plane Survey Extraordinaire (GLIMPSE~I; \citealt{benjamin03}) and the MIPS Inner Galactic Plane Survey (MIPSGAL; PI: Carey) using the InfraRed Array Camera (IRAC; 3.6 to 8.0\micronsns; \citealt{fazio04}) and the Multiband Imaging Photometer for Spitzer (MIPS; 24 to 160\micronsns; \citealt{rieke04}) respectively.

Both the MSX and GLIMPSE~I surveys include observations around 8\micronsns, and were taken $\sim$8 years apart. In this paper we make use of this similarity in wavelengths to search for sources with mid-IR variability between the MSX and the \emph{Spitzer} surveys. Our analysis is most sensitive to variability arising from dusty objects such as AGB stars or YSOs, and should be sensitive to sources with a wide range of variability timescales extending up to a decade.

In \S\ref{sec:observations}, we give a summary of the various datasets used for this study.
In \S\ref{sec:procedure}, we give an overview of the procedure used to select the sample of candidate variable stars.
In particular, we discuss the possible sources of contamination, i.e. sources that look - but are not - variable (\S\ref{sec:contaminants}); the issue of apparent variability due to strong spectral features and the differences in spectral responses (\S\ref{s:models}); and the details of the merging of the MSX and GLIMPSE~I catalogs (\S\ref{sec:bandmerging}). In \S\ref{s:sample} we compile a final list of candidate variable stars.
We then attempt to refine this sample to a subset of sources that present additional evidence for potential variability (\S\ref{s:refine}), and present Spectral Energy Distributions (SEDs) of the 40 sources in this subset, as well as images for three typical sources (\S\ref{sec:examples}).
Finally we discuss the possible nature of these sources (\S\ref{s:analysis}) and summarize our results (\S\ref{s:conclusion}).

\section{Observations}

\label{sec:observations}

In order to search for point source variability, we made use of the following data:
\begin{itemize}

\item The MSX survey of the Galactic plane \citep{price01}, which was carried out between 1996 and 1997, and consisted of imaging in four bands: A (8.28\micronsns),  C (12.13\micronsns),  D (14.65\micronsns), and E (21.3\micronsns).
\item The 2 Micron All-Sky Survey (2MASS; \citealt{skrutskie06}), which was carried out between 1997 and 2001. The observations consist of imaging in three bands: J (1.25\micronsns), H (1.65\micronsns), and \ks (2.15\micronsns).
\item The GLIMPSE~I survey of the Galactic mid-plane \citep{benjamin03}, which was carried out in 2004 at 3.6, 4.5, 5.8, and 8.0\microns using the IRAC instrument on board {\it Spitzer}
\item The MIPSGAL survey of the same region as GLIMPSE~I, carried out in 2005 and 2006 at 24, 70, and 160\microns using the MIPS instrument on board {\it Spitzer}. We make use only of the data taken at 24\micronsns.

\end{itemize}

The resolution of 2MASS and IRAC is $\sim$2\arcsec, that of MIPS is $\sim$10\arcsec, and the resolution of the MSX observations is $\sim$18\arcsec.

The approximate sensitivity and saturation levels for the various datasets are listed in Table~\ref{t:sens}. The values most critical for this work are the saturation level of \irac (1590mJy) and the sensitivity level of \msx (100 to 200mJy).

The absolute calibrations of the observations are described in detail in \citet{cohen_2mass}, \citet{reach05}, \citet{engelbracht07}, and \citet{price04}  for 2MASS, IRAC, \mipsns, and MSX respectively.

The spectral response curves of the instruments used for the various datasets are shown in Figure \ref{f:filters}. In this paper we make use of the similarity between the wavelength range covered by \irac and \msx (as well as the similarity between \mips and \msxe in \S\ref{s:refine}).

We used the GLIMPSE v2.0 Point-Source Catalog (Meade  et al., 2007\footnote{http://www.astro.wisc.edu/glimpse/glimpse1\_dataprod\_v2.0.pdf}), which includes IRAC fluxes for all the point sources in the survey, as well as 2MASS fluxes when available. We chose the Point Source Catalog over the Point Source Archive as it has a higher reliability and does not include saturated sources (for a more detailed description of the Catalogs and Archives, we refer the reader to Meade et al., 2007).
For MSX data, we used the MSX6C Point Source Catalog \citep{egan03}. We restricted our analysis to the ($\ell$,$b$) ranges covered by GLIMPSE I, i.e. $10^\circ < |\ell| < 65^\circ$ and $|b| < 1^\circ$. The total area covered is 220 square degrees.

\section{The selection of the variable star candidates}

\label{sec:procedure}

\subsection{Contaminants}

\label{sec:contaminants}

In order to search for variable stars in the mid-infrared, we would ideally use observations of the Galactic plane at two different epochs taken with the same telescope and with the same filters. However, in this case we have data for two surveys using \emph{similar} but not \emph{identical} spectral responses, namely \irac and \msxns, at spatial resolutions differing by almost a factor of 10, and sensitivity and saturation limits overlapping by only an order of magnitude. Therefore, great care has to be taken when comparing the two datasets.

There are three main sources of possible contamination, i.e. sources that could look variable between \irac and \msx based on a comparison of the images or catalogs:

\paragraph{Confusion} The difference in the resolutions between MSX and IRAC observations (18\arcsec\,and 2\arcsec\,respectively) means that multiple sources in \irac can appear as one source in \msxns. Therefore, if one were to merge an MSX source with the closest IRAC source, one might find that the fluxes appear to differ, when in fact this is because other nearby sources in IRAC also contribute to the flux of the MSX source.

\paragraph{Saturation} If a source is mildly saturated in \iracns. the response of the detector becomes nonlinear, and the source will appear fainter than it really is. Since MSX has a much lower sensitivity, many sources detected in MSX are saturated in GLIMPSE, so this effect could be common. Therefore, we choose to use the GLIMPSE Catalog, which does not include saturated sources. More generally, since we can use only very bright (non saturated) sources in GLIMPSE and faint sources in MSX, it is essential in this analysis to perform a rigorous quality check of all fluxes by inspecting the images to ensure there are no `false sources' or other technical problems linked to saturation or noise that could affect flux measurements.

\paragraph{Spectral features} Since \irac and \msx do not cover exactly the same wavelength range (as shown in Figure \ref{f:filters}), the two fluxes for a same source may differ if the spectrum contains strong spectral features in the wavelength range covered by the instruments. For example, YSOs with a strong  silicate feature at 10\microns in absorption (e.g. embedded protostars), or in emission (e.g. young stars with disks), are likely to be affected by this. This effect will also be present in stars with large amounts of interstellar extinction, which produces a silicate absorption feature at 10\micronsns. In addition, the presence of strong emission or absorption features in the wavelength ranges covered by only one or the other of the instruments would also lead to differences in the fluxes. For example, the 6-6.5\microns region is covered only by \msxns, and cool giants and supergiants are known to show absorption at 6.3\microns due to the H$_2$O absorption bands \citep{cohen95,tsuji01,tsuji03}.\\

These issues mean that it would be extremely difficult to produce a statistically {\it complete} sample of variable star candidates. Instead for this work we choose to prioritize {\it reliability} over completeness.

\subsection{Strong spectral features and apparent variability}

\label{s:models}

As mentioned in \S\ref{sec:contaminants}, strong spectral features in the wavelength range covered by the \irac and \msx bandpasses could cause the two fluxes to differ. In order to assess how important this effect is for YSOs, we use the publicly available grid of 200,000 YSO model SEDs from \citet{robitaille06}. We use the convolved IRAC and MSX fluxes for each model for a $\sim$18,000\,AU aperture, which corresponds to 18\arcsec~(the resolution of the MSX observations) at 1kpc (a typical distance to objects in the GLIMPSE survey). We note that the choice of aperture has very little effect for the analysis presented here.

\begin{figure}
\epsscale{0.9}
\plotone{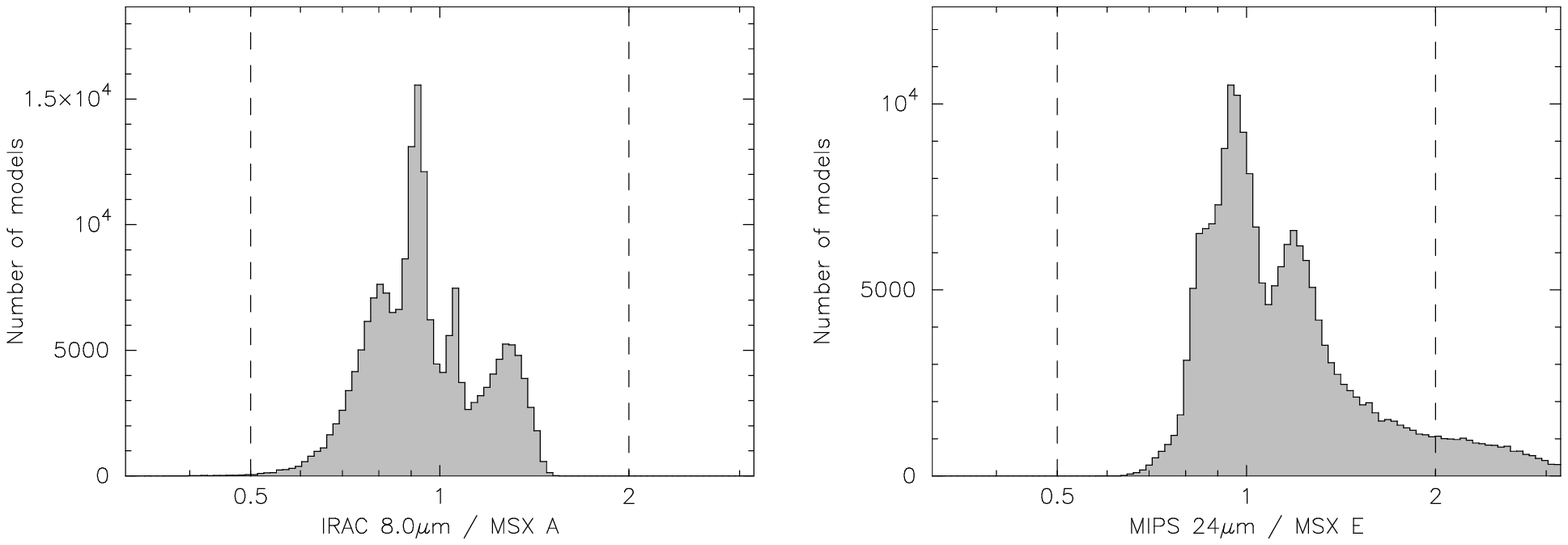}
\caption{Calculated ratios of the \irac fluxes to the \msx fluxes (left) and of the \mips fluxes to the \msxe fluxes (right) for the grid of YSO model SEDs presented in \citet{robitaille06}. The vertical lines delimit in each case the region within which the two fluxes agree to within a factor of two. The peaks in both distributions are artifacts due to the sampling of parameter space for the model grid (see \citeauthor{robitaille06} for more details on the sampling of the parameters).\label{f:ysomodels}}
\end{figure}

The left panel in Figure \ref{f:ysomodels} shows a histogram of the number of models as a function of the ratio of the \irac flux to the \msx flux for all the models.  Only for very few models (0.15\%) do the two fluxes differ by more than a factor of two. We note that although we use only YSO models for this analysis, results for AGB models are likely to be similar, since both types of objects show similar SEDs, i.e. thermal dust emission with a silicate feature.

We then investigated how much the ratio of \irac to \msx for a normal star would be affected by interstellar extinction. To do this we used a 4,000\,K model atmosphere from \citet{castelli04} with Log[g]=+2.0 and Log[Z/H]=-2.0, to which we applied extinctions of \av = 0, 20, 40, 60, 80, and 100 using the extinction law discussed in \citet{robitaille07}. We do not expect the interstellar extinction towards most sources to exceed this value, since \av~values of 100 or more are only seen through InfraRed Dark Clouds (IRDCs). Such sources would be easy to identify, since IRDCs are clearly visible against the background diffuse PAH emission: indeed, assuming $A_{[8.0]}/A_{K_{\rm s}}\approx0.5$ \citep{flaherty07} and $A_{K_{\rm s}}/A_{V}\approx0.1$ \citep{cardelli89}, \av=100 corresponds to $A_{[8.0]}\approx5$, i.e. a drop in brightness by a factor of 100. Only a small fraction of sources in the galactic mid-plane appear to be behind IRDCs, and most of the sources in our final sample of candidate variable stars do not appear to be situated close to any IRDCs. After applying the extinction, we convolved the synthetic spectra with the \irac and \msx spectral response curves (see Appendix~A of \citealt{robitaille07} for details of the convolution equations).

\begin{figure}
\epsscale{0.6}
\plotone{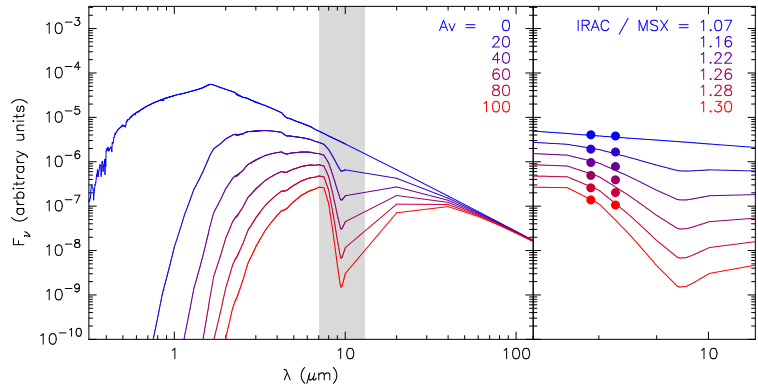}
\caption{The spectrum of a 4,000\,K star for values of the visual extinction \av~ranging from 0 to 100 (top to bottom spectra). The region indicated by the gray area in the left panel is shown in more detailed in the right panel. The points show the synthetic fluxes for \irac and \msxns. The ratio of the \irac to the \msx flux for each spectrum is given in the right panel.\label{f:models}}
\end{figure}

Figure \ref{f:models} show the six synthetic spectra, along with the convolved fluxes. The convolved \msx flux does not decrease as fast with increasing \av~ as the monochromatic flux at 8.28\micronsns. This is because as shown in Figure \ref{f:filters} the spectral response curve for \msx actually extends down to 6\microns (as does \iracns) and therefore a large fraction of the flux contributing to the in-band \msx flux does not originate from within the region affected by the silicate absorption feature. Even with \av=100, the ratio of \irac to \msx only reaches 1.3.

\subsection{Combining GLIMPSE~I and MSX}

\label{sec:bandmerging}

We extracted all the MSX point sources from the MSX6C Catalog for which an 8.28\microns flux was available. We kept only sources for which the uncertainty in the flux values at 8.28\microns was less than 10\% of the flux, and for which the reliability and confusion flags were set to 0 (indicating reliable data). The total number of sources satisfying these criteria and with coordinates in the GLIMPSE I area is 107,091.

In order to avoid the problem of confusion described in the previous section, we decided to use only MSX sources corresponding to a single GLIMPSE source. Specifically, we kept only MSX sources which had exactly one GLIMPSE source within 4\arcsec, and we further required the sum of the \irac flux of neighboring sources within 18\arcsec\,to not exceed 20\% of the flux of the central GLIMPSE source. In this way, we ensured that we only used MSX sources which corresponded to one GLIMPSE source, with a maximum confusion level of 20\%. This reduced the number of sources to consider from  107,091 to 62,013. Note that since the GLIMPSE~I Catalog does not include saturated sources as mentioned previously, we are not affected by the apparent variability arising from saturation.

Before looking for true variability, a further criterion was applied. It is possible for sources in \irac to not make it into the GLIMPSE~I Catalog, for example, saturated or extended sources, or even diffuse PAH emission. Therefore, even though there may not be bright neighbors \emph{in the Catalog} around a given source, we still need to check that the contamination level is low.

In order to do this, we modified the GLIMPSE~I mosaics to match the diffraction-limited and pixel resolutions of MSX: we first smoothed the images using a Gaussian filter so that stars had a half-width at half maximum of 18\arcsec; we then re-binned the mosaics by a factor of five to match the pixel resolution of 6\arcsec/pixel of MSX mosaics. The \msx mosaics and the \irac mosaics then look very similar. In fact, most stars are seen to have the same relative fluxes in both observations. We then performed PSF fitting photometry on these resampled GLIMPSE~I mosaics at the positions of the MSX sources selected above, and compared these fluxes to the fluxes of the matched GLIMPSE point sources found previously. If there was no contamination, one would expect the two fluxes to be very similar. This is the case for many - albeit not all - sources. For this study, we keep only sources for which the original and resampled \irac fluxes do not differ by more than 20\%.

Once these criteria were applied, we obtained a list of 50,744 sources. In summary, this list contains sources detected in \msxns, corresponding to a single unsaturated GLIMPSE source, with very little contamination ($<20$\%) due to confusion, whether from neighboring point sources or other sources of emission. We now use this sample of sources to search for variability.

\subsection{A list of variable star candidates}

\label{s:sample}

\begin{figure}
\epsscale{1.0}
\plotone{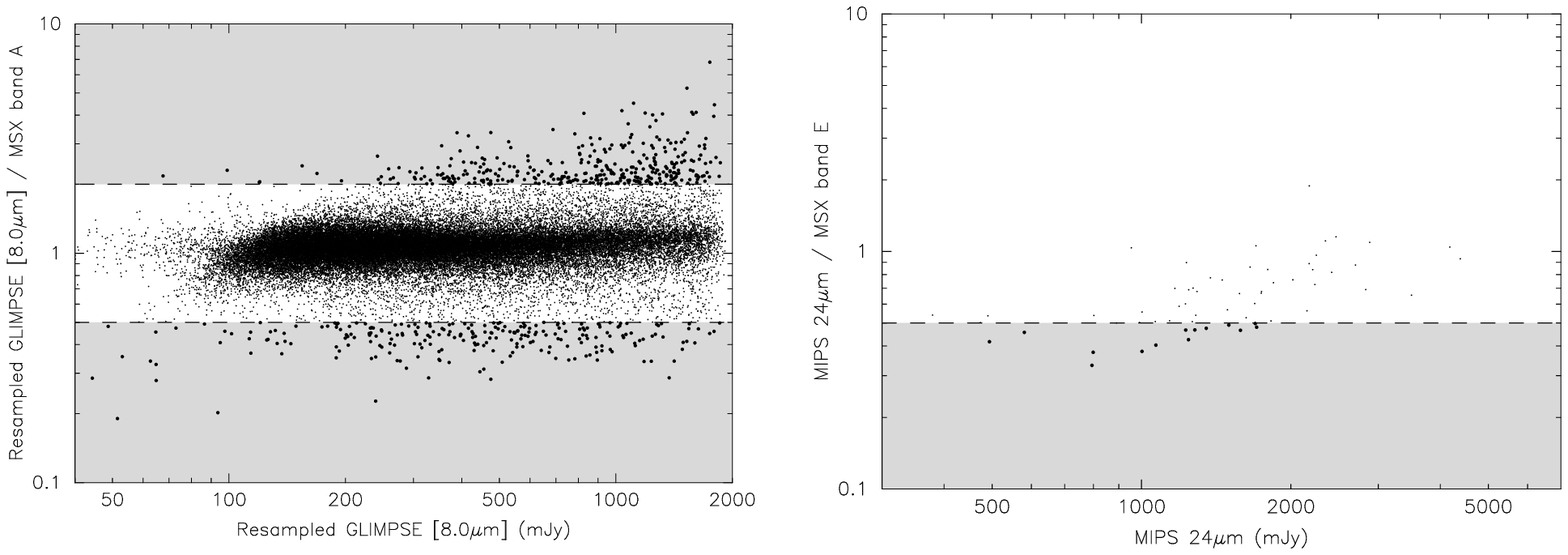}
\caption{Left: ratio of the \irac to the \msx fluxes for all sources for which the \irac flux from the GLIMPSE Catalog and from the smoothed mosaics agree to within 20\%.  The dashed lines indicate where the ratios are equal to 0.5 and 2, and the grey areas show the regions within which we flag sources as variable. Right: ratio of the \mips to the \msxe fluxes for all sources from the sample of 552 sources for which both fluxes were available. Similarly to the left panel, the dashed line indicates where the ratios are equal to 0.5, and the gray areas show the regions within which we flag sources as variable.\label{f:comparison}}
\end{figure}

In light of the results in \S\ref{s:models}, we decided to require the \irac flux from the smoothed GLIMPSE~I mosaics and the \msx flux to differ by more than a factor of two in order for a source to be considered as potentially variable. This cutoff is conservative since we are aiming for reliability rather than completeness. This reduces the number of sources in the sample from 50,744 to 592. Figure \ref{f:comparison} shows the \irac to \msx ratio of all 50,744 sources, with the variable star candidates highlighted. We visually inspected the \irac (both original and smoothed) images and residuals, as well as the \msx images, in order to check that the photometry was performed correctly in all cases. We eliminated 40 sources where the photometry at 8\microns on the smoothed mosaics was not reliable (due to confusion or extended emission), further reducing the sample to 552 sources.

These sources are listed in Table~\ref{t:sample}, along with their respective 2MASS, IRAC and MSX fluxes. In addition to inspecting the \irac and \msx images as mentioned above, we examined all the remaining MSX data (bands C, D, and E) using the standard (CB02) and high-sensitivity (CB03) MSX mosaics. We eliminated any unreliable fluxes where the corresponding source was very noisy or absent from the images, since we use the data in these bands for further source selection (\S\ref{s:refine}) and when plotting the SEDs. Therefore we can rule out bad data and unreliable photometry as a cause of apparent variability. We have also included \mips fluxes when data were available, as described in the next section.

Although we cannot rule out that strong spectral features around 10\microns or at 6-6.5\microns are causing offsets by more than a factor of two between \irac and \msxns, we see from the results in \S\ref{s:models} that such spectral variations cannot be explained by thermal dust spectrum with a silicate feature or a standard extinction law. Therefore, even if the sources in this sample are not all intrinsically variable, they remain very interesting objects for future follow-up studies.

\section{Highly likely variable star candidates}

\subsection{Source selection}

\label{s:refine}

In this section, we use the sample of 552 sources to search for a subset which present secondary indicators of variability. To do this, we use several criteria:

\paragraph{Very high variability}

We flag a source as highly likely to be variable if the fluxes at \irac and \msx differ by more than a factor of 4. It is extremely difficult to explain how such a flux could be due to strong spectral features, and we can rule out problems associated with the data, as we have visually quality-checked all the IRAC and MSX data used. Using this selection criterion, we find 11 sources.

\paragraph{JH\ksns/IRAC mismatch}

If a star is truly variable and presents a large difference (e.g. a factor of two difference) in flux at 8\micronsns, it is reasonable to assume that in some cases at least, the source should also be variable at shorter wavelengths. Since the 2MASS and IRAC data were taken a few years apart, we can therefore search for an apparent mismatch between JH\ks and IRAC fluxes. However, since these two wavelength ranges do not overlap, it is not straightforward to do this, and we restrict ourselves to cases where the mismatch is the most obvious.

We decided to remove sources with JH\ks and IRAC data consistent with the colors or stars or YSOs with no variability. This includes sources with colors consistent with those of AGB stars with no variability, since AGB stars are known to have  IRAC colors similar to those of YSOs (as can be seen by comparing the color-color diagrams in \citealt{robitaille06} and \citealt{marengo06}).

In order to do this, we use the SED fitting tool described in \citet{robitaille07}, to which we have added the grid of stellar photosphere models from \citet{castelli04}. The fitting tool uses linear regression to fit model stellar and YSO SEDs to observed SEDs, allowing the interstellar extinction and the distance to be free parameters.

We first fit all the sources, using only JH\ks and IRAC data, with the stellar photosphere models, and allowing a range of \av~ from 0 to 30. We then consider all sources fit with a $\chi^2$ per datapoint of less than 2 as being consistent with no mismatch. The cutoff value is arbitrary, but visual inspection shows that all fits with a smaller $\chi^2$ per datapoint are very good. We then fit the remaining sources with YSO models, again considering all sources with a $\chi^2$ per datapoint of less than 2 as being consistent with no mismatch. We assumed that the objects could lie anywhere from 300\,pc to 10\,kpc, with an interstellar \av~of up to 40. Out of the 552 sources in our list of variable star candidates, 496 had enough data to be used in the SED fitting tool (we required data for at least four wavelengths from JH\ks and IRAC); 78 were well fit by stellar photosphere models and 391 were well fit by YSO models, leaving 27 sources. These sources were in some cases badly fit for reasons other than a JH\ksns/IRAC mismatch, so after visual inspection of the SEDs we selected a sample of 15 sources for which an offset in the SED between JH\ks and IRAC was clearly the cause of the bad fit.

A mismatch between JH\ks and IRAC could be due to a high \ksns-band flux which can in turn be due to the confusion in the \ksns-band (this would occur more frequently at longitudes towards the inner galaxy). Therefore, we visually inspected the JH\ks and IRAC images for this subset of sources in order to ensure that this was not the case, and ruled this out for the 15 sources.

We can of course not rule out that the offset between 2MASS and IRAC is simply due to intrinsically exotic SEDs for these sources (e.g. due to near-IR spectral lines or multiple sources) but it seems likely that in most cases, variability is the main cause of offset, especially when considered alongside the fact that the \irac and \msx fluxes are also offset from each other by at least a factor of two.

\begin{figure}
\epsscale{1.0}
\plotone{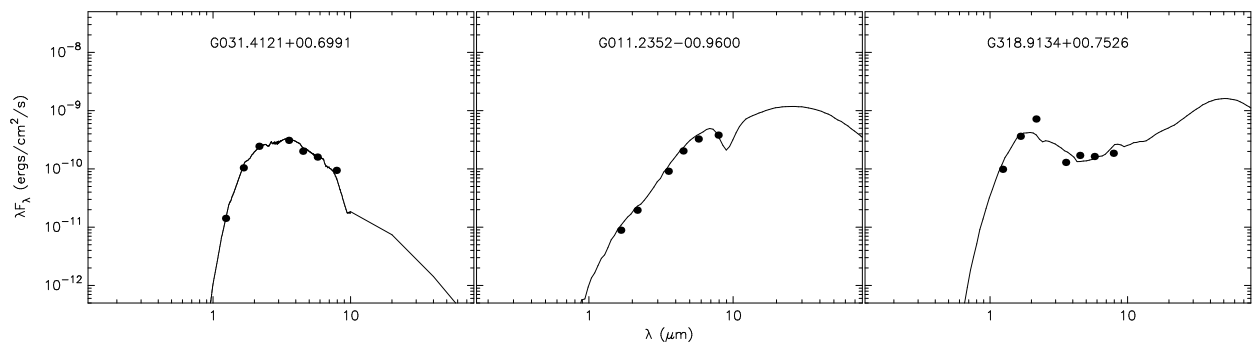}
\caption{Model SED fits to the broadband photometry of three example sources to find offsets between JH\ks and IRAC data. The data are shown as points and the models as solid lines. The error bars are smaller than the points. Left: a source whose data are consistent with a reddened stellar photosphere. Center: a source whose data are consistent with a model SED for an embedded protostar. Right: a source that is not well fit by model SEDs for stars or YSOs. The cause of the bad fit is the mismatch between the JH\ks and IRAC points. The model shown is the best fit obtained with YSO models\label{f:sedfits}.}
\end{figure}

In order to illustrate the above procedure, we show the SEDs of three sources in Figure~\ref{f:sedfits} with the best model SED fit in each case. These include a source well explained by a stellar photosphere model, a source well explained by a YSO model, and a source for which we could not fit any models well due to the offset between JH\ks and IRAC.

\paragraph{\msxe and \mips mismatch}

Finally, we flag sources as being likely variables if there is a clear offset between the \mips and the \msxe flux. We measured the \mips fluxes using PSF photometry on data available in the Spitzer Science Center Archive (MIPSGAL Program; PI Carey) for the 552 sources in the sample using a custom written PSF fitting program. Since we can ignore saturated pixels in the fitting, we are able to extract fluxes up to 2-3\,Jy for mildly saturated sources. As for \msx and \iracns, we have to be careful when looking for offsets between \mips and \msxe fluxes because of the difference in wavelength and spectral response (see Figure \ref{f:filters}). We performed an analysis - similar to that described in \S\ref{s:models} - for \msxe and \mipsns, and found that although ratios of \mipsns/\msxe larger than 2 are possible, due to steeply rising dust spectra at those wavelengths, ratios less than 0.5 cannot be reproduced by the models (see right-hand panel of Figure \ref{f:ysomodels}). Therefore, we select only sources where \mipsns/\msxe$ < 0.5$. As mentioned previously, the MSX images for all bands were checked visually, and the band E fluxes for the sources in our sample are reliable. Using this criterion, we find 14 sources. These are highlighted in the right-hand panel of Figure \ref{f:comparison}.

\begin{figure}
\plotone{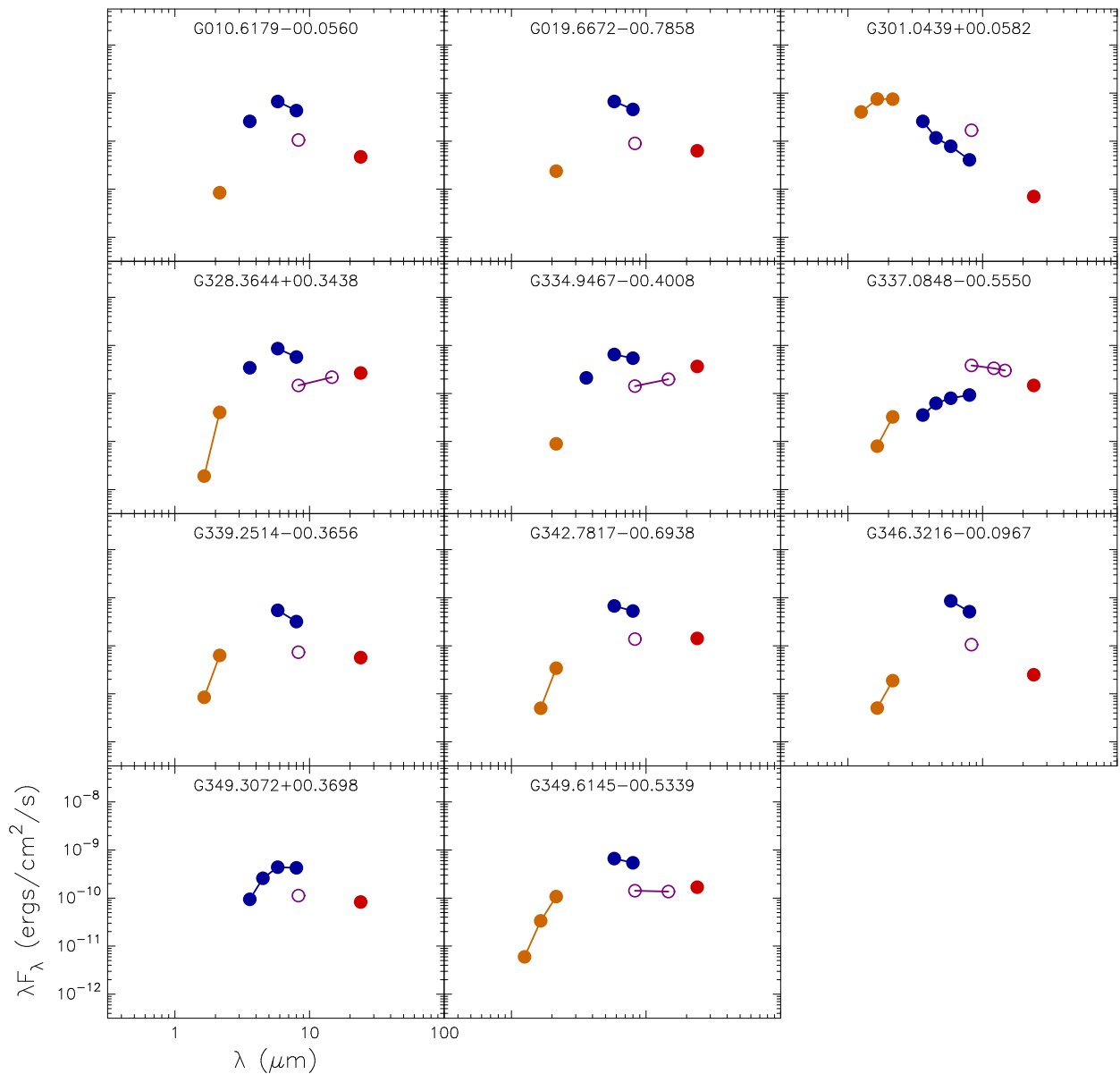}
\caption{The SEDs for the sources where \irac and \msx differ by more than a factor of four. The filled circles show (when available), from left to right, 2MASS, IRAC, and \mips data points. The open circles show (when available) the MSX data points. \label{f:seds1}}
\end{figure}

\begin{figure}
\plotone{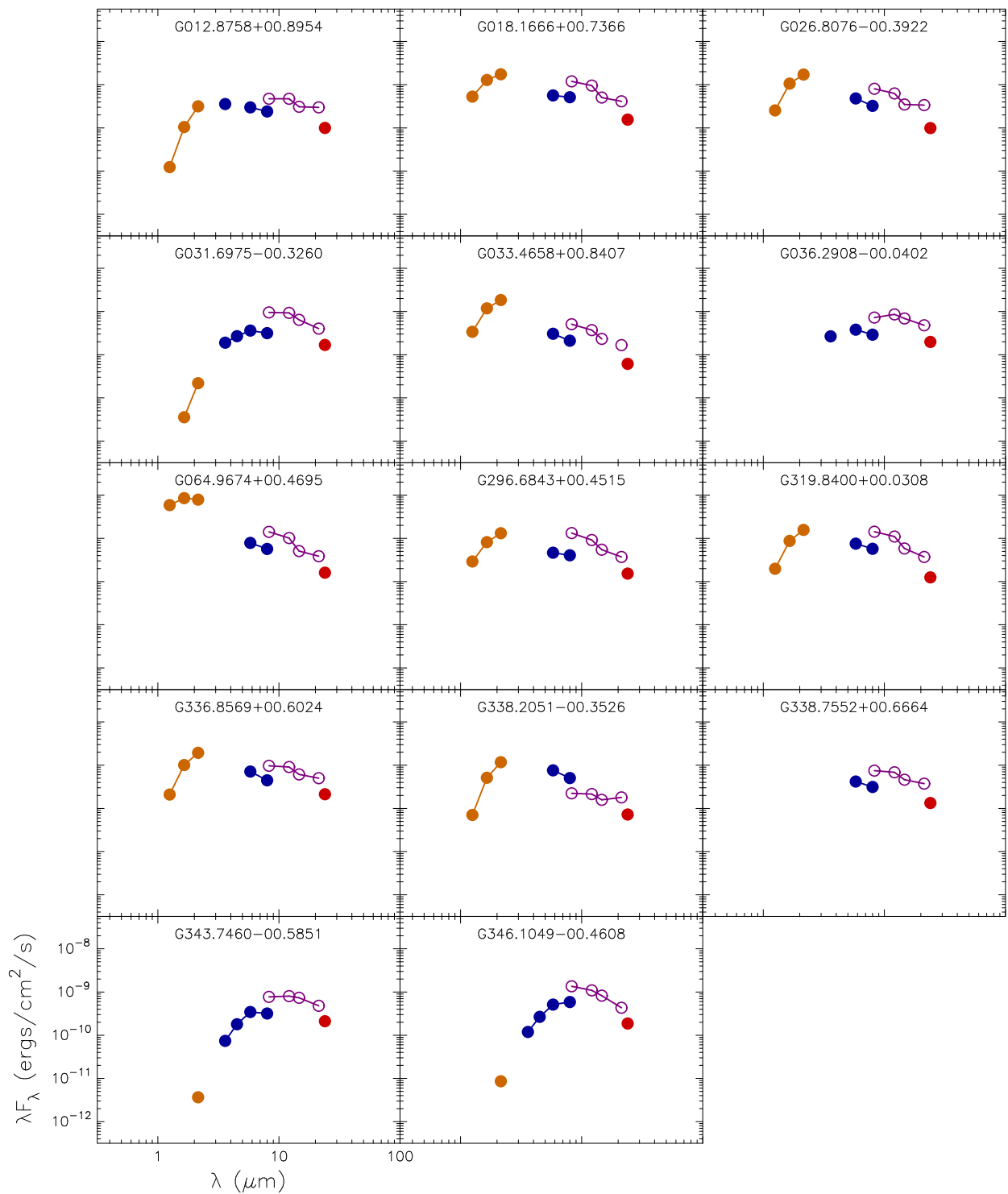}
\caption{The SEDs for the sources where in addition to the \irac and \msx fluxes differing by more than a factor of two, the \msxe and \mips fluxes also differ by more than a factor of two. The data points follow the same style as in Figure \ref{f:seds1}. \label{f:seds2}.}
\end{figure}

\begin{figure}
\plotone{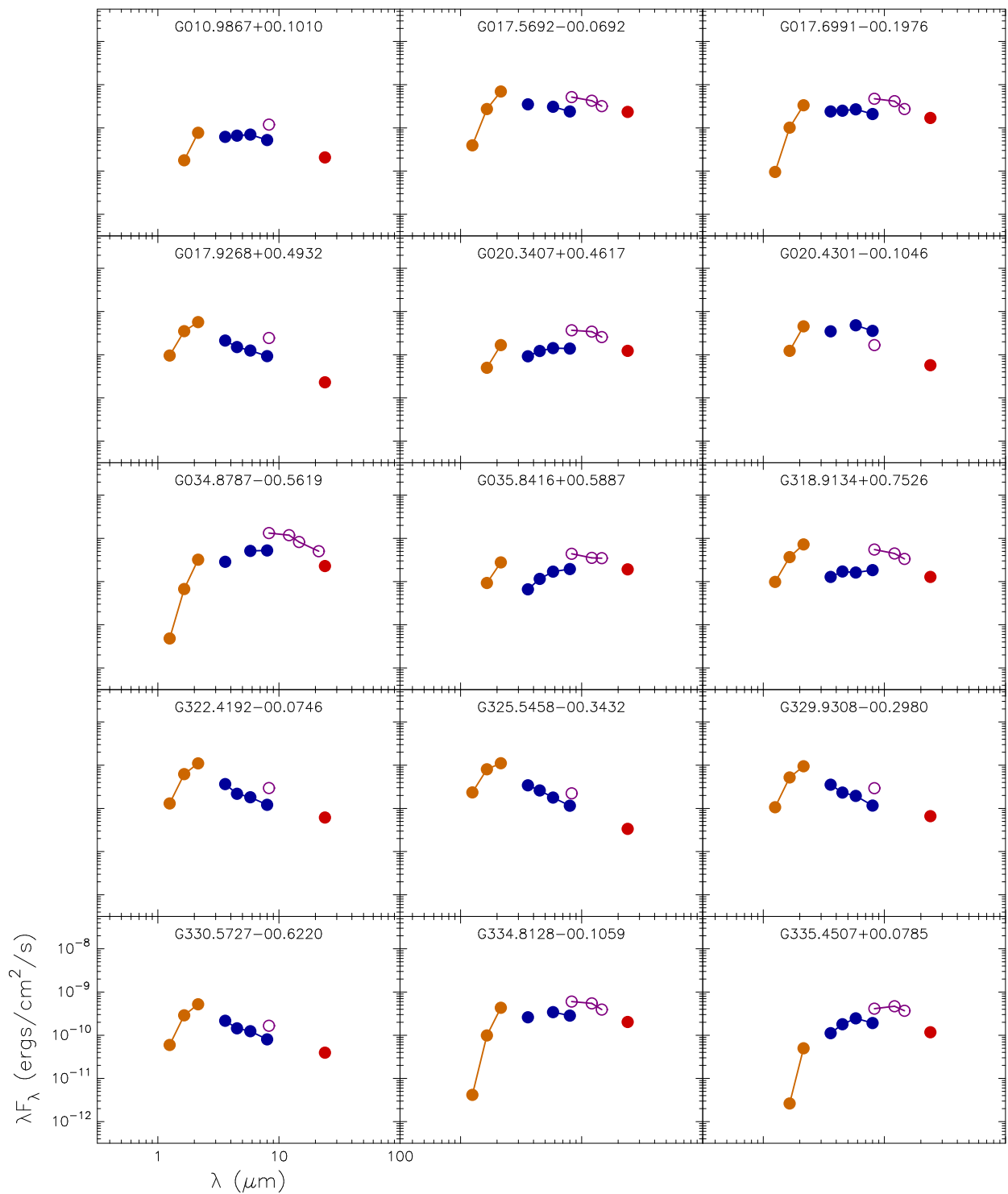}
\caption{The SEDs for the sources where in addition to the \irac and \msx fluxes differing by more than a factor of two, the JH\ks and IRAC data seem to be offset. The data points follow the same style as in Figure \ref{f:seds1}.\label{f:seds3}}
\end{figure}

Table \ref{t:subset} lists the 40 sources in our sample of 552 which satisfy at least one of the above three criteria. The SEDs for all 40 sources are presented in Figures \ref{f:seds1}, \ref{f:seds2}, and \ref{f:seds3}.

\subsection{Typical examples}

\label{sec:examples}

\begin{figure}
\epsscale{0.8}
\plotone{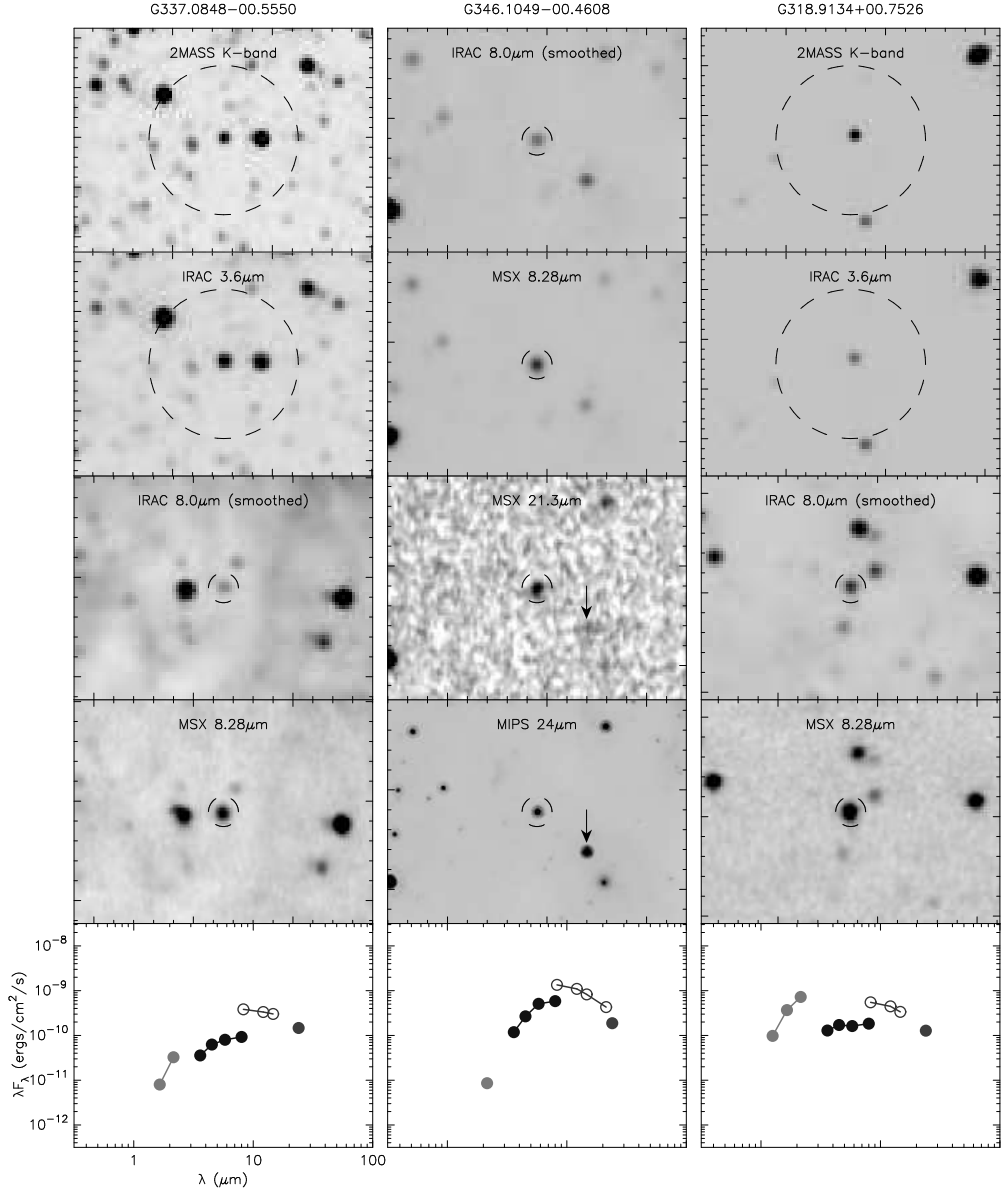}
\caption{Images for three sources. For each source, four images are shown, and the bottom panel shows the SED. The IRAC 3.6\microns images are slightly smoothed to match the 2MASS resolution. The circles in each plot are 27\arcsec~in radius (the K and IRAC 3.6\microns images are shown on a smaller scale). The arrow indicates G346.0803-00.4808, which did not make it to our sample of variable stars (as it is very mildly saturated in IRAC), but appears to be highly variable in \msxe and \mips (the flux differs by a factor of at least $\sim$2.5). The \ks and IRAC 3.6\microns plots for G318.9134$+$00.7526 are shown on a different stretch to show that the slope of the SED is bluer between \ks and IRAC 3.6\,$\mu$m than the surrounding stars. The three SEDs are presented in the same way as in Appendix A.\label{f:images}}
\end{figure}

In Figure \ref{f:images}, we show a selection of images for three sources, one from each of the above sub-samples in turn, as well as the SED for these sources. The sources are:

\paragraph{G337.0848$-$00.5550} This source (shown in the left panels) has a ratio of \irac to \msx of 0.23, and therefore is flagged as an `extreme variable'. The large variability is clearly seen by comparing the \irac and \msx images. This source also shows a possible JH\ks mismatch, but the offset was not large enough to make it through our selection process. However, in the light of the extreme variability in \iracns, it seems likely that the offset between JH\ks and IRAC wavelengths is real. There is no possible source of confusion at \ks and 3.6\micronsns, as shown in the images.

\paragraph{G346.1049$-$00.4608} The \mips and \msxe fluxes for this source (shown in the center panels) differ by more than a factor of two. The top two panels show the difference between the source brightness in \irac and \msxns, whereas the two panels below show the difference between \mips and \msxens. The difference is more difficult to see in the latter due to the high noise in the \msxe mosaic. However, whereas the central source is slightly brighter than the source shown in the top right in \msxens, the two sources have a more similar flux in \mipsns. A very interesting source is present to the bottom right - G346.0803$-$00.4808. It is brighter than G346.1049$-$00.4608 in \mipsns, yet is undetected in \msxens! Unfortunately it is not present in our sample of 552 sources, because its IRAC flux lies less than 5\% above the `mild saturation' limit.

\paragraph{G318.9134$+$00.7526} This source (shown in the right panels) shows a very large offset between JH\ks and IRAC, and this is clearly not due to any issues with the matching of the 2MASS to the GLIMPSE source after inspection of the images. In fact, the offset in the SED is so large that the slope from \ks to 3.6\microns is actually bluer than for a normal star. This can be seen in the two top panels - the central star actually becomes fainter relative to all the surrounding stars from \ks to 3.6\micronsns. As shown in the two panels below, the brightness of the source then increases significantly between \irac and \msxns. This source is probably our best candidate in the overall sample.

\section{Analysis and Discussion}

\label{s:analysis}

\subsection{Possible mechanisms of mid-IR variability}

\label{s:nature}

We now discuss the main mechanisms for variability in the mid-IR. The types of sources that are likely to show such variability are the following:

\paragraph{Asymptotic Giant Branch (AGB) stars} Long Period Variables (LPVs; classified into Mira, semiregular, and slow irregular variables), which are known to be AGB stars, show variability by factors of hundreds or thousands in the optical over timescales of months to years. These low- to intermediate-mass stars which are undergoing shell Helium fusion suffer large mass losses through stellar winds, and are thought to be one of the main dust production sites in the Galaxy. The dust chemistry can either be carbon or oxygen dominated, which leads to the two types of AGB stars being referred to as Carbon- or Oxygen-rich. A third type of AGB stars has been observed, with very red near-IR colors; these are referred to ``extreme'' AGB stars \citep[e.g.][]{blum06}, the brightest of which were previously identified as  ``obscured'' AGB stars \citep[e.g.][]{loup97}. These are mostly C-rich AGB stars with large amounts of circumstellar dust.

Optical to mid-IR studies of Mira variables have shown that the amplitude of the variability generally decreases with wavelength \citep{lockwood71,lebertre92,lebertre93,barthes96,smith02}, with the exception of the silicate feature around 10\micronsns, where the amplitude of the variations is higher than at immediately smaller and larger wavelengths \citep{lebertre93}. \citet{littlemarenin96} also show that the strength of the silicate feature in AU Cygni (an O-rich AGB star) varies with time. 

Key signatures of variability due to AGB stars are therefore the presence of a thermal IR excess, a decrease in the amplitude of the variations with wavelength, and a silicate feature with varying strength.

\paragraph{Pre-main-sequence stars} Optical and UV variability is common in pre-main-sequence stars. T Tauri stars for example are variable by definition \citep{joy45}; the variability in these objects has been attributed to hot and cool spots on the stellar surface, the former due to magnetospheric accretion, and the latter (also referred to as starspots) due to inhibition of the convection by a strong magnetic field. At longer wavelengths, where the bulk of the flux is due to thermal emission from the dust, variability is somewhat less common. However, various types of pre-main-sequence stars show infrared variability, including Herbig Ae/Be stars \citep{prusti94}, FU Ori type objects \citep{abraham04}, and UX Ori type objects \citep{juhasz07}.

Key signatures of pre-main-sequence stars include the presence of a thermal IR excess, and the presence of a silicate feature at 10 microns (whether in absorption or emission). These are a priori indistinguishable from the signatures expected for AGB stars. However, an additional  signature which is specific to YSOs would be the large-scale spatial correlation of variables in our sample with known star-formation regions.

\paragraph{Active Galactic Nuclei (AGNs)} Variability has been observed in AGNs on timescales of hours to years, and at X-ray, UV, optical, infrared, and radio wavelengths. In the mid- and far-infrared, evidence suggests that blazars, including BL Lac objects, are the most variable AGNs, while Seyfert 1 \& 2 galaxies and quasars are less variable \citep[e.g.][]{edelson87,sembay87}.

A key signature of AGNs would be a relatively uniform distribution in the plane of the sky, with no correlation with the distribution of Galactic sources.\\

\subsection{Physical nature of the variable star candidates}

\label{s:distribution}

\begin{figure}
\epsscale{1.0}
\plotone{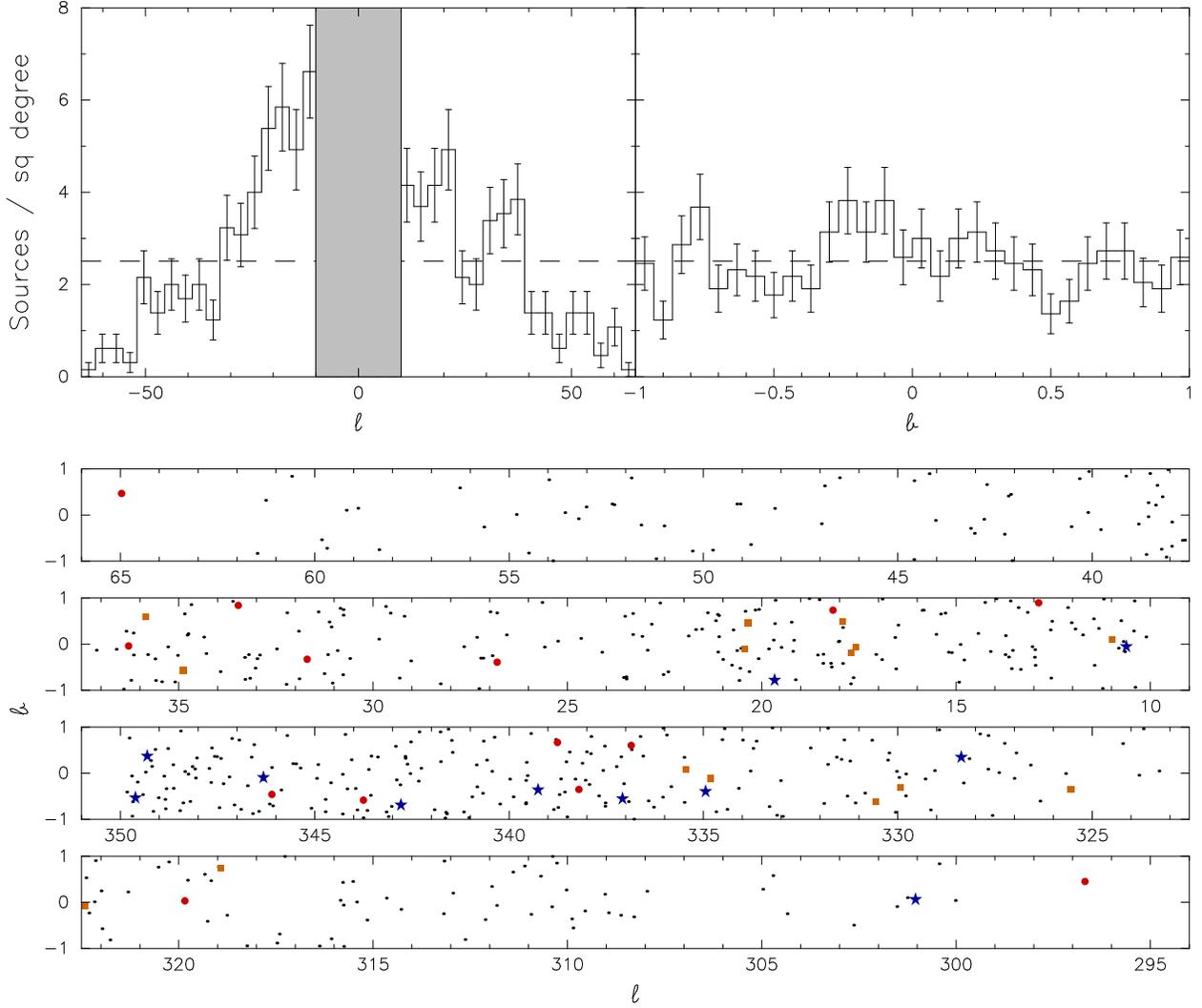}
\caption{Top: Galactic longitude and latitude distribution of the 552 candidate variable stars. The dashed line shows the expected level for a uniform distribution. The error bars were calculated using Poisson statistics. Bottom: Spatial distribution of the candidate variable stars (sources where \irac and \msx differ by more than a factor of 2). The dots represent sources in the whole sample of 552 sources. The stars show the location of the subset of sources where \irac and \msx differ by more than a factor of 4, the circles show the sources where \msxe and \mips differ by more than a factor of two, and the squares show the sources where a mismatch between JH\ks and IRAC is present.\label{f:spatial}}
\end{figure}

In order to determine whether the sources are most likely to be galactic (e.g. AGBs or YSOs) or extra-galactic (e.g. AGNs), we examine the spatial distribution of the sources in our list of variable star candidates. Figure \ref{f:spatial} shows the galactic longitude and latitude distribution of the 552 variable star candidates, as well as their spatial distribution. The number of variable sources clearly peaks towards the galactic center ($\sim6$/square degree at $|\ell|=10^\circ$), and tends to very low values ($<1$/square degree at $|\ell|=65^\circ$) further away from the center. This indicates that the large majority of the sources are galactic. If the sources were extragalactic, one should not observe any increase in source density towards the galactic center. In fact, one would expect a \emph{decrease} towards the galactic center due to the increased line-of-sight extinction through the galaxy. The latitude distribution is consistent with a flat distribution, but we note that this may simply be because we only extend out to $b=\pm1^\circ$

The Galactic distribution of the variable star candidates does not correlate strongly with the location of massive star formation regions  (e.g. M16 at $(\ell,b)\approx(15^\circ,-0.7^\circ)$, M17 at $(\ell,b)\approx(17^\circ,0.8^\circ)$, or the G305 complex at $(\ell,b)\approx(305.2^\circ,0.1^\circ)$), suggesting that the majority of sources are likely to be AGBs rather than YSOs.

Based purely on the Galactic longitude distribution of the sources, we can rule out extragalactic sources of variability such as AGNs for the vast majority of sources, and based on the spatial distribution, we favor the AGB interpretation.

\begin{figure}
\epsscale{0.30}
\plotone{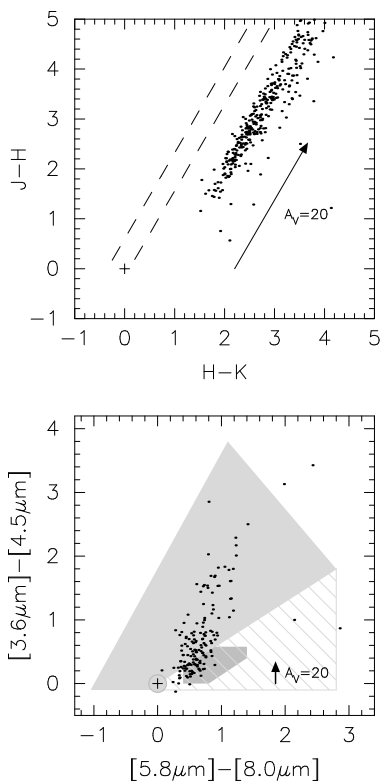}
\caption{JHK (top) and IRAC (bottom) color-color plots of the variable star candidates. Only sources with all three 2MASS fluxes were used for the JHK color-color plot, and only sources with all four IRAC bands were used for the IRAC color-color plot. The cross marks the approximate location of un-reddened stars, and the reddening vectors show an extinction of \av=20. The dashed lines in the top panel show where reddened stars would lie for a range of stellar temperatures. The filled and hashed ares in light gray in the bottom panel are adapted from \cite{robitaille06}, and show where YSOs are expected to lie in IRAC color-color space.\label{f:colorcolor}}
\end{figure}

In Figure \ref{f:colorcolor} we show the JH\ks colors of all the sources in the sample of 552 for which fluxes in J, H, and \ks bands were available, as well as the IRAC colors for all sources for which fluxes in all IRAC bands were available. The sources shown are clearly much redder than un-obscured stars. We show the locus for reddened photospheres in the JHK color-color plot, and the range of expected colors for YSOs in the IRAC color-color plot (adapted from \citealt{robitaille06}). Both the red H-K color and the IRAC colors of these sources suggests the presence of thermal dust emission, in agreement with both the YSO and the AGB interpretation.

Finally, we mention that a few sources in Figure \ref{f:seds3} show a wavelength-dependence for the variability. For example, G318.9134$+$00.7526 (also shown in Figure \ref{f:images}) and G035.8416$+$00.5887 both display a very large offset between \ks and IRAC 3.6\microns (at least an order of magnitude), whereas the offset at 8\microns is somewhat smaller (half an order of magnitude). This is what one would expect if these were AGB stars/Mira variables.

\subsection{Comparison to known sources}

\label{s:known}

We searched for the 552 sources from our list of variable star candidates in the Combined General Catalogue of Variable Stars (GCVS4.2; \citealt{samus04}) - which lists 430 sources in the GLIMPSE~I survey area - and found that none are listed as known or suspected variable stars. This is expected, as most of these are likely to be too extinguished to be seen at optical wavelengths.

One of the sources in the sub-sample of 40, G338.2051$-$00.3526, was found to match with a source in the ISOGAL survey classified as a YSO (ISOGAL-P J164144.9-465006; \citealt{felli02}).

We then compared our list of sources - before and after selection - to the IRAS Point-Source Catalog in order to verify whether our selection procedure increased the fraction of sources with high variability probability (the \texttt{Var} index). We searched for the 50,744 sources used for the initial selection of variable sources in the IRAS Point-Source Catalog, and found 3,292 with an IRAS source within 15\arcsec. Of the 552 variable candidates, 95 have a corresponding IRAS source, and of the 40 highly likely variable sources, 15 have a corresponding IRAS source. In Table \ref{t:flags} we list in each case the fraction of IRAS sources with a \texttt{Var} index of at least 95\%. We found that the initial selection of variable stars increased the fraction of IRAS sources with \texttt{Var}$>95$\% from 7.7\% to 27.4\%, and the further selection of candidates with a secondary indicator of variability increased the fraction further to 46.6\%.

We repeated this analysis by using the \texttt{var} flags from the MSX Point-Source Catalog (available for all sources, since all the sources in our lists originate from the MSX Point-Source Catalog). The results are also shown in Table \ref{t:flags}. The initial selection of variable sources increases the number of sources with the variability flag set in at least one band from 17.3\% to 42.8\%. When considering sources with at least two variability flags set, the fraction increases from 1.0\% to 6.3\% when selecting the 552 variable stars, and the refinement of the sample to 40 sources increases this to 12.5\%.

The above results indicate that one quarter to half of the sources in the list of 552 variable star candidates, and at least $40\%$ of the sub-sample of 40 are likely to be variable on timescales of months, i.e. the time between the multiple IRAS or MSX observations. If the remaining sources in our sample of 552 are truly variable, this suggests that the timescales of their variability has to be larger than the lifetime of the MSX and IRAS missions, and therefore has to be of the order of years. We have included both the \texttt{Var} index and $\sum$\texttt{var} (the sum of the variability flags) for each source in Table \ref{t:sample}.

\begin{figure}
\epsscale{1.0}
\plotone{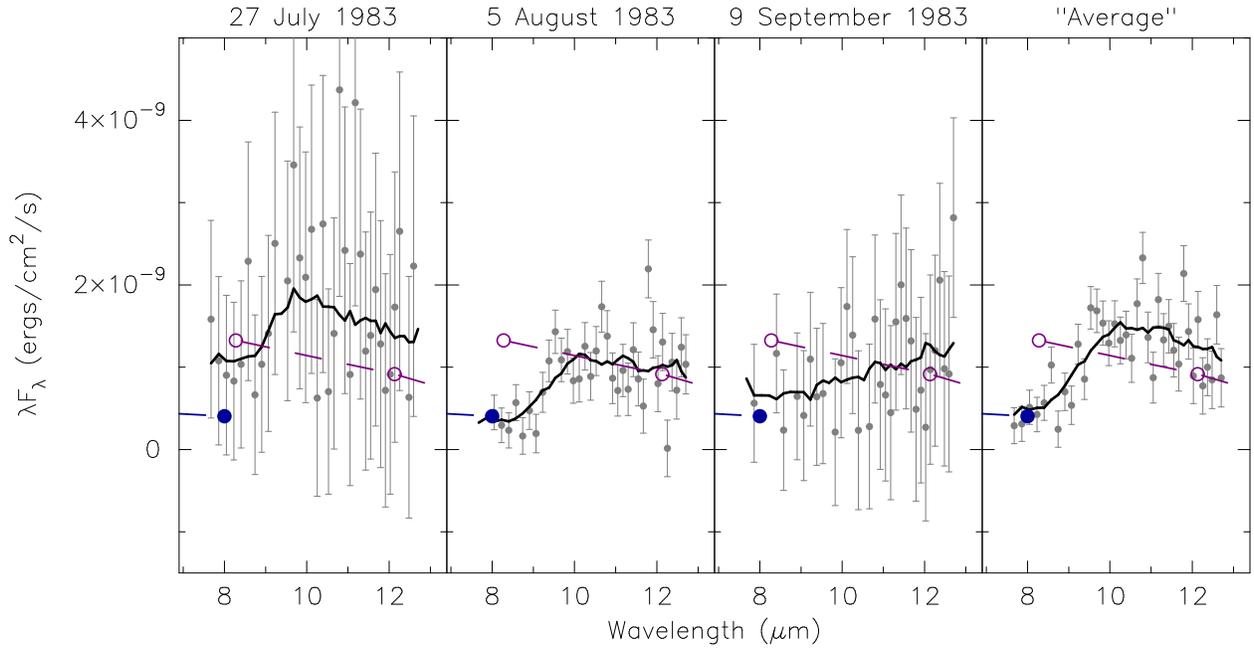}
\caption{LRS Spectrum for G296.6843+00.4515 (IRAS 11557-6129) for three epochs, and the ``Average'' LRS spectrum (as defined in the IRAS Explanatory Supplement, Chapter IX.C.3). The points and error bars show the original calibrated spectra (with negative fluxes removed), and the solid lines show the data after smoothing the data by performing a `box optimal average' at each point (the total width of the box was set to 1.5\micronsns). The larger open and closed circles, show the MSX and IRAC data respectively, and the dashed lines are equivalent to the solid lines in Figure \ref{f:seds2}.\label{f:lrs}}
\end{figure}

Finally, we searched the original Dutch Low Resolution Spectrograph (LRS) database in order to find mid-IR spectra for the sub-sample of 40 sources. Three sources were found to have LRS spectra, two of which were extremely noisy. The third, G296.6843+00.4515 (IRAS 11557-6129), shows a dominant broad 10\microns emission feature. Its width, peak, and shape all suggest the presence of O-rich dust in the form of amorphous silicates. This is shown in the right hand panel of Figure \ref{f:lrs}. The data shown are in fact the `Average'' \footnote{As defined in the IRAS Explanatory Supplement, Chapter IX.C.3} of three LRS spectra taken at different epochs, ranging over 43 days (27th July, 5th August, and 9th September 1983). We extracted the individual spectra for the three epochs, and have shown these in Figure \ref{f:lrs}. Despite the noise in the data, it does appear that: (1) the strength of the silicate feature changes over the three epochs, and (2) the continuum level also changes (as seen from the monochromatic flux at 8\microns). These signatures suggest that this source is indeed an O-rich AGB star displaying variability.

\section{Conclusion}

\label{s:conclusion}

We have selected a sample of 552 sources for which the \irac and \msx flux differ by more than a factor of two. We have checked that the data are of sufficient quality through selection criteria and visual inspection of the data for all these sources. We have ruled out that differences of more than a factor of two can be due to the silicate feature in a standard interstellar extinction law, or silicates seen in absorption or emission in YSOs using standard models. Therefore, this sample contains objects which are all potentially interesting for follow-up observations, even if not variable, as they may show a deviation from standard extinction, different emission processes in YSOs (such as PAH emission), or other strong spectral features in YSOs and AGB stars. If most of these sources are confirmed as variable, they will significantly expand the number of known variable stars in the Galactic plane: as mentioned in \S\ref{s:known}, the GCVS v4.2 only lists 430 sources in the GLIMPSE~I survey area.

We further refined this sample to a subset of sources for which a secondary indicator of variability was available. We selected sources for which the variability at 8\microns is very strong, with \irac and \msx differing by more than a factor of four. We also selected sources for which the \mips to \msxe ratio was less than half, and finally we selected sources for which the JH\ks and IRAC data appear to be offset. The number of sources in each category is 11, 14, and 15 respectively.

By comparing the lists of sources before and after selection with the \texttt{Var} index in the IRAS Point-Source Catalog and the \texttt{var} flags in the MSX Point-Source Catalog, we estimate that at least one quarter to half of sources in the sample of 552 sources are truly variable, on timescales up to months. The remaining sources are either truly variable, albeit on longer timescales (years), or contain strong spectral features which cause the offset between the \irac and the \msx flux.

We have also presented multi-epoch IRAS LRS spectra which tentatively confirm variability in one of the sources (G296.6843+00.4515), and show a silicate feature varying in strength over time, suggesting that this sources is an O-rich AGB star.

The analysis we have performed is a preliminary selection of the most prominent candidate variables, and by no means a complete census of mid-IR variability in the Galaxy. We are biased towards bright sources in GLIMPSE and faint sources in MSX. Since we require sources to be unsaturated in GLIMPSE and detected in MSX, we are biased towards objects with redder spectral slopes. In addition, when searching for \msxe and \mips mismatches, or JH\ks and IRAC offsets, we are biasing the sample further towards objects with redder slopes, because the sensitivity in \msxe is even lower than \msxns. Despite all these restrictions, we find 552 sources with \irac and \msx fluxes differing by at least a factor of two, and 40 sources with secondary evidence for variability.

While we were restricted to using $\sim50,000$ sources that were in common between GLIMPSE~I and MSX and not affected by confusion, the total number of Catalog sources in the GLIMPSE~I survey is around 31,000,000. This suggests that our sample is really just a very small tip of the Galactic iceberg, and that the true number of variable sources is very likely to be of the order of tens of thousands.

\acknowledgments

We wish to thank the anonymous referee, and Bob Benjamin, for useful suggestions.
This research made use of data products from the Midcourse Space 
Experiment; the NASA/IPAC Infrared Science Archive, which is operated by the 
Jet Propulsion Laboratory, California Institute of Technology, 
under contract with the National Aeronautics and Space 
Administration; and the Vizier service \citep{ochsenbein00}.
Partial support for this work was provided by a Scottish Universities Physics Alliance Studentship (TR).
MC thanks NASA for supporting his participation in this work through contract 1242593 with the University of California, Berkeley.
Additional support, as part of the Spitzer Space Telescope Theoretical Research Program (BW, TR), Legacy Science Program (EC, MM, BB, BW), and Fellowship Program (RI), was provided by NASA through contracts issued by the Jet Propulsion Laboratory, California Institute of Technology under a contract with NASA.

%\bibliographystyle{apj}
%\bibliography{apj-jour,references} 
\bibliography{}

\clearpage

\begin{figure}
\epsscale{1.0}
\plotone{f1.eps}
\caption{The spectral response curves for 2MASS JH\ksns, {\it Spitzer} IRAC and \mipsns, and MSX. Note that there is significant overlap between the \irac and \msx filters, and between the \msxe and \mips filters. \label{f:filters}}
\end{figure}

\clearpage

\clearpage

\clearpage

\clearpage

\clearpage

\clearpage

\clearpage

\clearpage

\clearpage

\clearpage

\clearpage

\begin{deluxetable}{lcc}
\tabletypesize{\scriptsize}
\tablecolumns{3}
\tablewidth{0pt}
\tablecaption{Sensitivity and saturation levels\label{t:sens}}
\tablehead{Band & Sensitivity & Saturation \\ & mJy & mJy}
\startdata
J & 0.4 & 16000\\
H & 0.6 & 16000\\
\ksns & 0.8 & 16000\\
IRAC 3.6\micronsns & 0.5 & 439 \\
IRAC 4.5\micronsns & 0.5 & 450 \\
IRAC 5.8\micronsns & 2.0 & 2930 \\
IRAC 8.0\micronsns & 5.0 & 1590 \\
MSX A & 100$\rightarrow$200& n/a\tablenotemark{a} \\
MSX C & 1100$\rightarrow$3100 & n/a\tablenotemark{a}\\
MSX D & 900$\rightarrow$2000 & n/a\tablenotemark{a}\\
MSX E & 2000$\rightarrow$6000& n/a\tablenotemark{a}\\
MIPS 24\micronsns & 2.6 & 1400 \\
\enddata
\tablenotetext{a}{According to the Midcourse Space Experiment Point Source Catalog Version 1.2 Explanatory Guide, only four sources in the whole survey were saturated. The saturation limit is likely to be above $10^6$\,mJy.}
\tablecomments{The above values are approximate, since they depend in some cases on factors such as confusion or background emission, and only apply to the surveys used in this paper. The 2MASS values are from the 2MASS All-Sky Data Release Explanatory Supplement (I.6), the IRAC values are from the GLIMPSE data products description, and the MSX sensitivity values are from the Midcourse Space Experiment Point Source Catalog Version 1.2 Explanatory Guide (Table 1). The \mips values are from a private communication.}
\end{deluxetable}

\clearpage

\pagestyle{empty}
\setlength{\voffset}{2cm}

\begin{deluxetable}{llrrrrrrrrrrrrrrrr}
\rotate
\tabletypesize{\tiny}
\tablecolumns{8}
\tablewidth{0pt}
\tablecaption{The sample of 552 sources with apparent variability at 8\microns \label{t:sample}}

\tablehead{ & & \multicolumn{2}{c}{Coordinates} & \multicolumn{3}{c}{2MASS} & \multicolumn{4}{c}{IRAC} & \multicolumn{4}{c}{MSX} & MIPS & MSX $\sum$\texttt{var}\tablenotemark{a} & IRAS \texttt{Var}  \\
MSX source name & GLIMPSE source name & $\alpha$ (J2000) & $\delta$ (J2000) & F$_{\rm J}$& F$_{\rm H}$& F$_{\rm K_{\rm s}}$ & F$_{3.6}$ & F$_{4.5}$ & F$_{5.8}$ & F$_{8.0}$& F$_{8.28}$ & F$_{12.13}$ & F$_{14.65}$ & F$_{21.3}$ & F$_{24}$ & & \\
& & $^\circ$ & $^\circ$ & mJy  & mJy  & mJy  & mJy  & mJy  & mJy  & mJy  & mJy  & mJy  & mJy  & mJy  & mJy & & \% }
\startdata
G010.1167$+$00.1543 & SSTGLMC G010.1162$+$00.1539 & 271.85730 & $-$20.11289 &   \nodata &   \nodata &    2.8 &  191.6 &   \nodata &  938.2 &  976.6 &  340.9 &   \nodata &   \nodata &   \nodata &  683.3 & 0 & 29 \\
G010.3957$-$00.9373 & SSTGLMC G010.3955$-$00.9365 & 273.01970 & $-$20.39660 &   14.3 &  109.4 &  297.0 &   \nodata &   \nodata &  447.3 &  402.9 &  187.6 &   \nodata &   \nodata &   \nodata &  339.5 & 0 & \nodata \\
G010.4174$+$00.9357 & SSTGLMC G010.4167$+$00.9355 & 271.28840 & $-$19.46910 &    2.2 &    3.2 &    5.1 &    5.5 &    7.8 &    7.2 &   55.6 &  149.8 &   \nodata &  690.0 & 2139.9 & 2466.1 & 0 & $-$1 \\
G010.4355$+$00.2827 & SSTGLMC G010.4351$+$00.2830 & 271.90300 & $-$19.77190 &    0.5 &   13.9 &  100.6 &   \nodata &   \nodata &  647.1 &  537.6 &  215.7 &   \nodata &   \nodata &   \nodata &  564.3 & 1 & \nodata \\
G010.6179$-$00.0560 & SSTGLMC G010.6180$-$00.0559 & 272.31180 & $-$19.77690 &   \nodata &   \nodata &    6.1 &  309.8 &   \nodata & 1290.0 & 1149.0 &  291.3 &   \nodata &   \nodata &   \nodata &  378.0 & 1 & 13 \\
G010.6616$-$00.1658 & SSTGLMC G010.6622$-$00.1653 & 272.43640 & $-$19.79180 &   \nodata &    5.7 &  147.1 &   \nodata &   \nodata & 1189.0 &  591.5 &  222.1 &   \nodata &   \nodata &   \nodata &  142.4 & 1 & \nodata \\
G010.7026$+$00.0569 & SSTGLMC G010.7020$+$00.0568 & 272.25040 & $-$19.64800 &   \nodata &    7.1 &   96.0 &  410.4 &   \nodata &  590.1 &  448.5 &  178.6 &   \nodata &   \nodata &   \nodata &  160.4 & 0 & \nodata \\
G010.7615$+$00.5088 & SSTGLMC G010.7611$+$00.5091 & 271.86160 & $-$19.37710 &   10.3 &   94.0 &  293.2 &  416.6 &  414.2 &  469.1 &  497.7 & 1216.3 & 1665.2 & 1156.4 &   \nodata &  895.5 & 1 & \nodata \\
\enddata
\tablecomments{The uncertainties are not shown for clarity. However, these are bright sources at mid-IR wavelengths in the {\it Spitzer} data, and therefore the uncertainties are very low in IRAC and \mipsns. In fact, the main source of uncertainty at IRAC and \mips wavelengths are likely to be the calibration uncertainties. For MSX data, one of the requirements used in Section \ref{sec:bandmerging} was the uncertainty at 8.28\microns should be less than 10\%. For the JHK and MSX errors we refer the reader to the original catalogs. The \irac flux calculated from the smoothed mosaics is not shown, but for sources to be in this list we required the original \irac flux to agree with our calculated \irac flux from smoothed mosaics by less than 20\%. Finally, we note that this table does not show all fluxes present in the original MSX Catalog (i.e. fluxes have been removed when a source was not seen, or seen with a very low signal-to-noise, in the images).}
\end{deluxetable}

\clearpage

\pagestyle{plaintop}
\setlength{\voffset}{0cm}

\begin{deluxetable}{lccccc}
\tabletypesize{\scriptsize}
\tablecolumns{6}
\tablewidth{0pt}
\tablecaption{Comparison to previously suspected variables\label{t:flags}}
\tablehead{ & \multicolumn{3}{c}{MSX sources} & \multicolumn{2}{c}{IRAS sources (matched)} \\
 & All & $\sum$\texttt{var}$\ge$1\tablenotemark{a} & $\sum$\texttt{var}$\ge$2\tablenotemark{a} & All & \texttt{Var}$>$95\% }
\startdata
Initial source list & 50744 & 8783 (17.3\%) & 504 (1.0\%) & 2695 & 207 (7.7\%) \\
Variable candidates & 552 & 236 (42.8\%) & 35 (6.3\%) & 95 & 26 (27.4\%) \\
Highly likely candidates & 40 & 17 (42.5\%) & 5 (12.5\%) & 15 & 7 (46.7\%) \\
\enddata
\tablenotetext{a}{We define $\sum$\texttt{var}=\texttt{vara+varc+vard+vare}, i.e. the sum of the variability flags for bands A, C, D, and E from the MSX Point-Source Catalog}
\end{deluxetable}

\clearpage

\begin{deluxetable}{ll}

\tabletypesize{\small}
\tablecolumns{2}
\tablewidth{0pt}
\tablecaption{The 40 sources with secondary indicators of variability\label{t:subset}}
\tablehead{MSX source name & GLIMPSE source name}
\startdata
\multicolumn{2}{c}{\bf Large 8\microns variability} \\
G010.6179$-$00.0560 & SSTGLMC G010.6180$-$00.0559 \\
G019.6672$-$00.7858 & SSTGLMC G019.6675$-$00.7870 \\
G301.0439$+$00.0582 & SSTGLMC G301.0440$+$00.0576 \\
G328.3644$+$00.3438 & SSTGLMC G328.3647$+$00.3434 \\
G334.9467$-$00.4008 & SSTGLMC G334.9466$-$00.4005 \\
G337.0848$-$00.5550 & SSTGLMC G337.0848$-$00.5550 \\
G339.2514$-$00.3656 & SSTGLMC G339.2511$-$00.3650 \\
\enddata
\end{deluxetable}

\end{document}